\UseRawInputEncoding
\documentclass[transaction]{IEEEtran}

\usepackage{cite}
\usepackage[cmex10]{amsmath}
\usepackage{graphicx}
\usepackage{picinpar}
\usepackage{amsmath,amsfonts,amssymb}
\usepackage{array}
\usepackage{mdwmath}
\usepackage{mdwtab}
\usepackage{eqparbox}
\usepackage{stfloats}
\usepackage{subfigure}
\usepackage{ragged2e}
\usepackage{extarrows}
\usepackage{booktabs}
\usepackage{multirow}
\usepackage{etoolbox}
\usepackage{makecell}
\usepackage[justification=centering]{caption}

\usepackage{algorithm}
\usepackage{algorithmic}
\usepackage{fixltx2e}
\usepackage{xcolor}
\usepackage{bm}

\begin{document}
\newtheorem{lemma}{Lemma}
\newtheorem{corol}{Corollary}
\newtheorem{theorem}{Theorem}
\newtheorem{proposition}{Proposition}
\newtheorem{definition}{Definition}
\newcommand{\e}{\begin{equation}}
\newcommand{\ee}{\end{equation}}
\newcommand{\eqn}{\begin{eqnarray}}
\newcommand{\eeqn}{\end{eqnarray}}
\renewcommand{\algorithmicrequire}{ \textbf{Input:}} 
\renewcommand{\algorithmicensure}{ \textbf{Output:}} 
\renewcommand{\raggedright}{\leftskip=0pt \rightskip=0pt plus 0cm}

\title{Massive Access in Extra Large-Scale MIMO with Mixed-ADC over Near-Field Channels\vspace{-0mm}}

\author{
Yikun Mei, Zhen Gao, De Mi,~\IEEEmembership{Member,~IEEE}, Mingyu Zhou, Dezhi Zheng, Michail Matthaiou,~\IEEEmembership{Senior Member,~IEEE}, Pei Xiao,~\IEEEmembership{Senior Member,~IEEE}, and Robert Schober,~\IEEEmembership{Fellow,~IEEE}
\vspace{-9mm}

\thanks{This work was supported in part by the Natural Science Foundation of China (NSFC) under Grant 62071044 and Grant U2001210; in part by the Shandong Province Natural Science Foundation under Grant ZR2022YQ62; in part by the Beijing Nova Program. \emph{(Corresponding authors: Zhen Gao; Mingyu Zhou.)}}

\thanks{Y. Mei is with the School of Information and Electronics, Beijing Institute of Technology, Beijing 100081, China.
Z. Gao and D. Zheng are with the MIIT Key Laboratory of Complex-Field Intelligent Sensing, Beijing Institute of Technology, Beijing 100081, China, also with the Yangtze Delta Region Academy, Beijing Institute of Technology (Jiaxing), Jiaxing 314019, China, and also with the Advanced Technology Research Institute, Beijing Institute of Technology, Jinan 250307, China.
D. M and P. Xiao are with University of Surrey, Guildford GU2 7XH, U.K.
M. Zhou is with Baicells Technologies Co. Ltd., Beijing 100089, China.
M. Matthaiou is with Queen’s University Belfast, Belfast BT3 9DT, U.K.
R. Schober is with Friedrich-Alexander Universit\"{a}t Erlangen-N\"{u}rnberg, 91054 Erlangen, Germany. (e-mails: gaozhen16@bit.edu.cn; zhoumingyu@baicells.com)}
}
%
\maketitle

\begin{abstract}
Massive connectivity for extra large-scale multi-input multi-output (XL-MIMO) systems is a challenging issue due to the near-field access channels and the prohibitive cost. In this paper, we propose an uplink grant-free massive access scheme for XL-MIMO systems, in which a mixed-analog-to-digital converters (ADC) architecture is adopted to strike the right balance between access performance and power consumption. By exploiting the spatial-domain structured sparsity and the piecewise angular-domain cluster sparsity of massive access channels, a compressive sensing (CS)-based two-stage orthogonal approximate message passing algorithm is proposed to efficiently solve the joint activity detection and channel estimation problem. Particularly, high-precision quantized measurements are leveraged to perform accurate hyper-parameter estimation, thereby facilitating the activity detection. Moreover, we adopt a subarray-wise estimation strategy to overcome the severe angular-domain energy dispersion problem which is caused by the near-field effect in XL-MIMO channels. Simulation results verify the superiority of our proposed algorithm over state-of-the-art CS algorithms for massive access based on XL-MIMO with mixed-ADC architectures. 
\end{abstract}


\IEEEpeerreviewmaketitle

\vspace{-6mm}
\section{Introduction}
%
Massive machine-type communication (mMTC) has long been identified as a major enabler of the Internet-of-Everything that will seamlessly connect a vast number of devices in future wireless networks{\cite{WP_6G, CXM_JSAC, ZJH_CC}}. To achieve the low latency, high efficiency, and reliability required for mMTC, uplink grant-free massive access schemes leveraging non-orthogonal radio resources have been widely investigated in recent years {\cite{CXM_JSAC, KML_TSP, KML_JSAC, SXD_TWC, CW_TSP, MYK_TWC}}. Due to the inherent sporadic traffic of mMTC, compressive sensing (CS)-based solutions have been proposed for joint activity detection and channel estimation (JADCE) yielding good performance {\cite{KML_TSP, KML_JSAC, SXD_TWC, CW_TSP}}. In \cite{KML_TSP}, an approximate message passing (AMP)-based detection algorithm was proposed to solve the JADCE problem by leveraging the virtual angular domain sparsity of massive multi-input multi-output (MIMO) channels. Furthermore, the authors of \cite{KML_JSAC} explored cooperation-based massive access for cell-free massive MIMO systems to avoid inter-cell interference. In the context of millimeter-wave/terahertz massive MIMO systems, the authors of {\cite{SXD_TWC}} proposed two multi-rank aware JADCE algorithms to exploit the simultaneously sparse and low-rank structure of the delay-angular domain channels. 

However, it is believed that massive MIMO-based 5G networks will gradually reach their limits. To accommodate denser connections with higher data rates in 6G networks, deploying extra large-scale MIMO (XL-MIMO) at the base station (BS) with extremely large array aperture has recently been proposed as a promising 6G candidate technology \cite{DLL_CL, DLL_Tcom}. Compared with massive MIMO, the emerging XL-MIMO substantially increases the near-field region, causing the access devices to fall into the near-field electromagnetic propagation environment. In this case, the near-field channels should be carefully modeled using spherical wavefronts, rather than the conventional far-field plane-wave assumption, requiring dedicated signal processing methods. To this end, near-field channel estimation has been investigated for XL-MIMO systems in {\cite{DLL_CL, JS_WCL, DLL_Tcom}} by considering grant-based multiple access while supporting a relatively small number of users. State-of-the-art JADCE solutions for massive access are based on the far-field plane-wave assumption, and their performance will degrade in near-field MIMO channels. 
Additionally, it is unrealistic to deploy high-resolution analog-to-digital converters (ADC) across all the antennas of XL-MIMO systems. To reduce cost and power consumption, MIMO receivers with low-resolution ADCs have been proposed in {\cite{Heath_TSP,JS_TSP}}. 
 However, the quantization noise caused by low-resolution ADCs significantly reduces the accuracy of the received signals, limiting detection performance.
To improve performance, mixed-ADC architectures utilizing a large number of low-resolution ADCs with a small number of high-resolution ADCs were investigated in \cite{ZW_JSAC,JS_TWC}, and the optimal ADC bit allocation for massive MIMO systems was analyzed in \cite{MM_TWC}.

In this paper, we design a CS-based grant-free random access scheme for massive connectivity based on XL-MIMO employing the mixed-ADC architecture, where a two-stage orthogonal approximate message passing (TS-OAMP) algorithm is proposed to solve the JADCE problem. Specifically, the common support structure of the spatial-domain channel and the cluster sparsity of the piecewise angular-domain channel are exploited to enhance the performance of activity detection and channel estimation, respectively. Moreover, a small number of high-resolution ADCs can ensure accurate hyper-parameter estimation by using the expectation-maximization (EM) algorithm, and the adopted subarray-wise estimation is capable of mitigating the angular-domain energy dispersion of near-field MIMO channels. Simulation results demonstrate that the proposed algorithm outperforms the state-of-the-art JADCE schemes for massive access in XL-MIMO systems.

\emph{Notations:} Column vectors and matrices are denoted by boldface lower and upper-case symbols, respectively; $(\cdot)^{\rm T}$, $(\cdot)^{\rm H}$ and $(\cdot)^{-1}$ denote the transpose, conjugate transpose, and inverse of a matrix, respectively; $\|\cdot\|_p$ is the $\ell_p$ norm of the argument; ${\bf A}_{:,n}$ denotes the $n$-th column vector of matrix $\bf A$; $\Re \{\cdot\}$ and $\Im\{\cdot\}$ denote the real and imaginary parts of the argument, respectively; 
$\text{diag}(\bf a)$ and $\text{diag}(\bf A)$ denote a diagonal matrix with $\bf a$ as the main diagonal and the diagonal vector of matrix $\bf A$, respectively;
$\text{Bdiag}(\cdot)$ denotes a block diagonal matrix created from input matrices; ${\cal C}{\cal N}({\bm{\mu}},{\bf \Sigma})$ denotes the complex Gaussian distribution with mean vector $\bm{\mu}$ and covariance matrix ${\bf \Sigma}$; $\Phi(\cdot)$ and $\phi(\cdot)$ are the cumulative distribution function and probability distribution function of a standard Gaussian random variable, respectively; ${\rm sign}(\cdot)$ and $\delta(\cdot)$ denote the signum function and the Dirac function, respectively;
$|\cdot|_c$ denotes the the cardinality of a set.

\vspace{-2mm}
\section{System Model}
We consider uplink grant-free massive Internet-of-Things (IoT) access based on XL-MIMO systems. The BS deploys an extra large-scale uniform linear array (ULA) comprising $N_r$ antenna elements, where the distance between two adjacent antenna elements is half a wavelength. There are $K$ single-antenna IoT devices in the coverage area of the BS. Due to the inherent sporadic traffic in mMTC, though $K$ can be very large, the number of simultaneously active devices $K_a$ can be small $(K_a \ll K)$. During the training phase, the $k$-th active device transmits the unique pilot sequence ${{\bf{s}}_k}{\in \mathbb{C}^{M}}$ to the BS, and the received signal ${{\bf{Y}}}{\in \mathbb{C}^{M \times N_r}}$ can be expressed as
\vspace{-2mm}
\begin{equation}\label{ysum}
{{\bf{Y}}}  = \sum\limits_{k = 1}^K {{\alpha _{k}}}{\sqrt{P_k}}{{\bf{s}}_k}{{\bf{h}}^{\rm T}_{k}} + {{\bf{W}}} = {\bf{S}}{{\bf{H}}} + {{\bf{W}}},
\vspace{-1mm}
\end{equation}
where ${{\alpha _{k}}}$ is the binary activity indicator that equals one when the $k$-th device is active and zero otherwise, ${\bf{h}}_{k}{\in \mathbb{C}^{N_r}}$ is the channel vector between the $k$-th device and the extra large array at the BS, $P_k$ is the $k$-th device's transmit power, ${{\bf{W}}}$ is the additive white Gaussian noise, each column of which follows ${\cal CN}{\left({\bf 0},\sigma^2 {\bf I}_M\right)}$, ${\bf{S}} = \left[{{\bf{s}}_1},\ldots,{{\bf{s}}_K}\right]{\in \mathbb{C}^{M \times K}}$, and ${{\bf{H}}} = \left[{{\alpha _{1}}}{\sqrt{P_1}}{{\bf{h}}_{1}}, \ldots,{{\alpha _{K}}}{\sqrt{P_K}}{{\bf{h}}_{K}}\right]^{\rm T}{\in \mathbb{C}^{K \times N_r}}$. The pilot matrix $\bf S$ is generated by randomly selecting $M$ rows from a $K$-dimensional discrete Fourier transform (DFT) matrix and satisfies ${\bf S}{\bf S}^{\rm H} = {\bf{I}}_M$.

The extra-large array aperture of the BS effectively increases the Rayleigh distance, which determines the boundary between the far and near fields. For instance, assuming $N_r = 512$ and a carrier wavelength of $\lambda = 0.05$ m, the theoretical Rayleigh distance is around $6.5$ km \cite{DLL_Tcom}. Thus, the near-field region is large and will cover most IoT devices around the BS. This illustrates that XL-MIMO systems significantly increase the extent of the near-field region. Thus, in typical XL-MIMO systems, most IoT devices are located within the near-field coverage, and the channels should be modeled based on spherical wavefronts instead of the conventional planar wavefronts.
In this case, the narrowband channel between the $n$-th antenna and the $k$-th device can be expressed as ${{{h}}_{k,n}} = \beta_{k,n} {{{\tilde h}}_{k,n}}$, where $\beta_{k,n}$ and ${{{\tilde h}}_{k,n}}$ are the large-scale fading and the small-scale fading coefficients, respectively. 
Moreover, ${{{\tilde h}}_{k,n}}$ can be modeled as
\vspace{-2mm}
\begin{equation} \label{Chmodel}
 {{{\tilde h}}_{k,n}} = \sqrt{\frac{\kappa_k}{\kappa_k+1}}e^{j\frac{2\pi}{\lambda}d^{\rm LoS}_{k,n}} + \sqrt{\frac{1}{\kappa_k+1}}\sum\limits_{l = 1}^{L_p}e^{j\frac{2\pi }{\lambda}d^{\rm NLoS}_{k,n,l}},
\vspace{-1mm}
\end{equation}
where $\kappa_k$ is the Rician factor for the $k$-th device, $\lambda$ is the wavelength, $d^{\rm LoS}_{k,n}$ and $d^{\rm NLoS}_{k,n,l}$ are the distances of the line of sight (LoS) component and the $l$-th non-LoS (NLoS) component between the $n$-th antenna and the $k$-th device, respectively, and $L_p$ is the number of NLoS components.

%

\begin{figure}[!tp]
	\centering
	\vspace{-2mm}
	\includegraphics[width=4.5 cm, keepaspectratio]
	{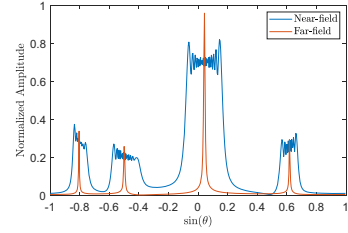}
	\captionsetup{font={footnotesize}, singlelinecheck = off, justification = raggedright,name={Fig.},labelsep=period}
	\caption{Comparison between the near-field and far-field angular-domain channels, where $\theta$ represents the AoA,  the distance of the LoS component in the near-field and far-field environments is $48$ m and $1000$ m, respectively, $N_r = 512$, and $L_p = 3$.}
	\label{fig1}
	\vspace{-5mm}
\end{figure}

Note that conventional far-field channels can be modeled based on the 3GPP spatial channel model (SCM) \cite{3GPP_SCM}, which assumes that for each multipath component, all receive antennas at the BS share the same angle-of-arrival (AoA). However, in XL-MIMO systems, this assumption may not hold, since the distance between the receive antennas and the devices can vary dramatically for different receive antennas. This \textit{near-field effect} in XL-MIMO channels destroys the Fourier transform relationship between the spatial-domain channel and the angular-domain channel that exists for far-field propagation, leading to significant energy dispersion, as shown in Fig. \ref{fig1}. Correspondingly, the traditional far-field massive MIMO channel estimation methods exploiting angular-domain sparsity are not applicable in near-field XL-MIMO systems. As a consequence, it is necessary to develop dedicated massive access approaches for near-field XL-MIMO systems.


\vspace{-2mm}
\section{Proposed CS-Based JADCE Algorithm}
During one frame, the active devices send their pilot sequence to the BS in the training phase. Once the identity information and the channel state information of the active devices are obtained, the BS can recover the data transmitted by these active devices through existing techniques \cite{JS_JSTSP}. Consequently, we focus on how to efficiently solve the JADCE problem.

In view of the high cost of configuring high-resolution ADCs for all antennas, we consider a mixed-ADC architecture. Before the quantization, a variable gain amplifier (VGA) with automatic gain control (AGC) is applied to scale the analog signal to the appropriate range. Then, the quantized received signal ${\widetilde{\bf Y}}$ can be expressed as
\vspace{-2mm}
\begin{equation}\label{quantization}
	{\widetilde{\bf Y}}_{:,n} = g\left({\bf{Y}}_{:,n}\right) =\begin{cases}
		{\bf{Y}}_{:,n},&n \in {\cal H}\\
		{\cal Q}_B\left({{\bf{Y}}}_{:,n}\right),&n \in {\cal L}
	\end{cases}
    ,\ \ \ \forall n,
\vspace{-1mm}
\end{equation}
where ${{\cal Q}_B}(\cdot)$ is the complex $B$-bit quantizer, ${\cal H}$ and ${\cal L}$ are the index sets of the RF chains employing high-resolution and low-resolution ADCs, respectively. In practice, the quantizer is applied to the received signals element-wise, with the real and imaginary parts quantized separately. For simplicity, here we consider uniform quantization. Define a set of thresholds $r_0 < r_1 < r_2< \cdots < r_{2^B}$, where $r_0 = -\infty, r_{2^B} = \infty, r_b = -1 + 2^{1-B}b, b = 1, 2, \cdots,2^B-1 $. For any input element $Y \in (r_{b-1}, r_b]$, ${\widetilde Y} = -1 + 2^{-B}(2b-1)$  is the output after quantization.

\begin{figure}[!t]
	\centering
	\vspace{-6mm}
	\includegraphics[width=7 cm, keepaspectratio]
	{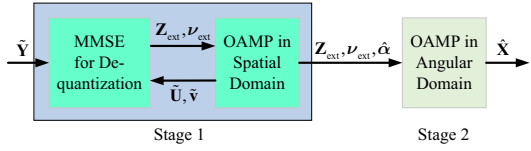}
	\captionsetup{font={footnotesize}, singlelinecheck = off, justification = raggedright,name={Fig.},labelsep=period}
	\caption{Block diagram of the proposed TS-OAMP algorithm.}
	\label{blockdigram}
	\vspace*{-5mm}
\end{figure}

To leverage the sparsity of the angular-domain channel, the quantized signals ${\widetilde{\bf Y}}$ need to be multiplied by the dictionary matrix. However, directly processing ${\widetilde{\bf Y}}$ using the conventional JADCE approaches \cite{KML_JSAC,KML_TSP,SXD_TWC} would not harness the gains of high-resolution quantized measurements, since the proportion of high-precision quantized measurements is small. On the other hand, it has been demonstrated that the common sparsity of the spatial-domain channel on different antennas can be exploited to enhance activity detection performance \cite{KML_TSP}. 
Inspired by these considerations, we propose a TS-OAMP algorithm to solve the JADCE problem with the generalized linear model (GLM) in (\ref{quantization}). As shown in Fig. \ref{blockdigram}, the proposed algorithm can be divided into two stages as follows:

$\bullet$ \textbf{Stage 1}: This stage identifies the active devices based on the problem formulation of the spatial-domain channel with common support sparsity. Moreover, the noise variance is estimated through the EM algorithm using high-resolution quantized signals.

$\bullet$ \textbf{Stage 2}: By leveraging the intermediate results in \textit{Stage} 1, this stage aims to estimate the piecewise angular-domain channel with cluster sparsity. To avoid the significant energy spread caused by the near-field effect, we decompose the whole array into several subarrays, and each subarray is individually processed based on a conventional far-field analysis.  

\subsubsection{Stage 1}
A similar problem for the GLM caused by the limited quantization has been solved in \cite{KML_JSAC, JS_TSP, JS_TWC}. Inspired by these works, we define two modules: a non-linear module and a standard linear model (SLM) module. For simplicity, we first define ${\bf{Z}} = {\bf{SH}}$, so that the quantized received signal is given by
${\widetilde{\bf Y}} = g\left({\bf{Z}} + {{\bf{W}}}\right)$.
%
%

The non-linear module performs de-quantization based on the minimum mean square error (MMSE) estimator and outputs the extrinsic messages as the input to the SLM module. Since the quantization function is non-linear, the element-wise MMSE estimator is employed to obtain the posterior mean $\hat{\bf Z}$ and variance ${\bm \nu}$, reducing the strong quantization noise. After that, by denoting the extrinsic message of the non-linear module as ${\bf{Z}}_{\rm ext}$ and ${{\bm{\nu}}_{\rm ext}}$, the equivalent SLM problem can be written as
\vspace{-2mm}
\begin{equation}\label{y_SLM_space}
	{\bf{Z}}_{\rm ext} = {\bf{S}}{{\bf{H}}} + {{\bf{W}_{\rm ext}}},
\vspace{-1mm}
\end{equation}
where the $n$-th column of the equivalent noise ${\bf{W}_{\rm ext}}$ is characterized by ${\cal CN}{\left({\bf 0},\nu_{{\rm ext},n}{\bf I}_M\right)}$. Note that ${{\bf{H}}}$ is the spatial-domain channel and exhibits the common sparsity structure, namely
\vspace{-2mm}
\begin{equation}\label{common_sparsity}
	{\rm supp}\{{\bf{h}}_{n}\}  = {\rm supp}\{{\bf{h}}_{n'}\} , \ \ \forall n,n' \in {\cal H}, n\neq n',
\vspace{-1mm}
\end{equation}
where ${{\bf{h}}}_{n}={{\bf{H}}}_{:,n}$ and ${\rm supp}\{\cdot\}$ denotes the set of non-zero elements of a vector.

To solve the typical multiple measurement vector (MMV) CS problem in (\ref{y_SLM_space}), we adopt the OAMP-MMV algorithm in \cite{MYK_TWC}. Generally, the OAMP algorithm proposed in \cite{OAMP} is more robust than the traditional AMP algorithm and has fewer requirements on the sensing matrix. Specifically, the OAMP algorithm consists of a de-correlated linear estimator and a divergence-free estimator, and the final output is the MMSE estimate when the algorithm converges. Furthermore, the OAMP-MMV algorithm was developed in \cite{MYK_TWC} to solve the MMV CS problem with a common sparsity structure. Based on this Bayesian framework, we assume that the \textit{a priori} distribution of $h_{k,n}$ is
\vspace{-1mm}
\begin{equation}\label{priori}
	p\left( {{h_{k,n}}} \right) = \left( {1 - {\lambda _{k,n}}} \right)\delta \left( {{h_{k,n}}} \right) + {{{\lambda _{k,n}}}}{\cal CN}{\left(0,\psi_{n}\right)},
\vspace{-1mm}
\end{equation}
%
where the \textit{a priori} sparsity ratio $\lambda_{k,n}$ denotes the non-zero probability of $h_{k,n}$. Then, the posterior mean and variance are obtained with the MMSE estimator of the OAMP-MMV algorithm\footnotemark[1]. \footnotetext[1]{Due to space constraints, we omit the details of the derivation of the OAMP-MMV algorithm and refer interested readers to \cite{MYK_TWC} for more information.}Subsequently, the SLM module passes extrinsic messages to the input of the non-linear module. 

The steps of \textit{Stage} 1 are summarized in lines 1-24 of {\textbf{Algorithm 1}}, where $T_1$ is the maximum number of iterations for this stage. The non-linear module comprises lines 4-11 for de-quantization, where ${\bf \hat Z}$ and ${\bm \nu}$ are calculated in lines 4-10 with some abuse of notation, since they are applied to the real and imaginary parts, respectively. The extrinsic message is given in line 11, where high-resolution and low-resolution quantized signals are handled differently. This is intuitive because the signals with high-resolution quantization are accurate. Lines 12-22 describe the OAMP-MMV algorithm, which constitutes the SLM module. The posterior sparsity ratio ${\pi}_{k,n}$ is given in line 15, which represents the posterior non-zero probability of $h_{k,n}$. Lines 19-21 are the EM steps to update the unknown parameters. In line 20, the common sparsity structure is exploited to update the sparsity ratio ${\lambda}_{k,n}$ by averaging, and ${\cal M}$ is defined as
\vspace{-1mm}
\begin{equation}\label{activity_selection}
	{\cal M} = \begin{cases}
		\left\{1,2,...,N_r\right\},&(\sigma^2)^{t-1} \geq (\sigma^2)_{\text{Th}},\\
		{\cal H},&\text{otherwise},
	\end{cases}
\vspace{-1.5mm}
\end{equation}
where the threshold $(\sigma^2)_{\text{Th}} = \frac{1}{|{\cal H}|_c}\sum\limits_{n\in {\cal H}} \frac{\|\tilde{\bf Y}_{:,n}\|_2^2}{(\text{SNR}_{\text{Th}}+1)M}$ to distinguish low signal-to-noise ratio (SNR) levels and high SNR levels is obtained according to a coarse estimate of the noise variance in Eq. (35) of \cite{KML_TSP}, and $\text{SNR}_{\text{Th}}$ is set to $10$ based on our computer experiments. In (\ref{activity_selection}), a small number of high-resolution signals are insufficient to perform accurate activity detection at low SNR levels. Thus, low-resolution signals are exploited to enhance the common sparsity structure. However, they should be removed from activity detection when the SNR is high, avoiding the interference caused by quantization noise. Moreover, the channel variance $\psi_n$ and noise variance $\sigma^2$ are updated as lines 19 and 21, respectively, where we also use the high-resolution signals to update the noise variance. On 
the contrary, previous works either assume that the noise
\begin{algorithm}[H]
	\caption{Proposed TS-OAMP Algorithm}\label{Algorithm1}
	\begin{algorithmic}[1]
		\vspace{-0.6mm}
		\REQUIRE {${\widetilde{\bf Y}}$, ${\bf{S}}$, and ${\bf{D}}_{\mathrm{sub}}$.}
		\ENSURE {${\hat{{\bf{X}}}}$ and $\hat{\bm{\alpha}}$.}
		\STATE {\textbf{\%\textit{Stage} 1}}:
		\STATE Initialize ${\tilde{\bf{u}}_n^0 = \bm{0}, {\bf{u}}_n^0 = \bm{0}}$, and ${{{\tilde v}_n^0}} =1$. $({\sigma^2})^{0}$, $\psi_n^0$, and ${{\lambda^{0}_{k,n}}}$ are initialized as Eqs. (35)-(37) of \cite{KML_TSP};
		\FOR {$ t=1$ {\textbf{to}} $ T_1$}
		
		\STATE $\mathring{y}_{m,n} = \Re\{{\tilde y}_{m,n}\} \in (r_{b-1}, r_b], \mathring{u}^{t-1}_{m,n} = \Re\{{\tilde u}^{t-1}_{m,n}\}$;   
		\STATE $\eta^t_{m,n} = \frac{{\rm sign}(\mathring{y}_{m,n})\mathring{u}^{t-1}_{m,n}-{\rm min}\{|r_{b-1}|,|r_b|\}}{\sqrt{[(\sigma^2)^{t-1}+{\tilde v}^{t-1}_n]/2}}$;
		\STATE $\xi^t_{m,n} = \frac{{\rm sign}(\mathring{y}_{m,n})\mathring{u}^{t-1}_{m,n}-{\rm max}\{|r_{b-1}|,|r_b|\}}{\sqrt{[(\sigma^2)^{t-1}+{\tilde v}^{t-1}_n]/2}}$;
		
		\STATE \hspace{-1.5mm}${\hat z}_{m,n,\Re}^t = \mathring{u}^{t-1}_{m,n} + \frac{{\rm sign}(\mathring{y}_{m,n}){\tilde v}_n^{t-1}}{\sqrt{2[{(\sigma^2)^{t-1}}+{\tilde v}_n^{t-1}]}}$$ \left[\frac{\phi(\eta^t_{m,n})-\phi(\xi^t_{m,n})}{\Phi(\eta^t_{m,n})-\Phi(\xi^t_{m,n})}\right]$;
		\STATE ${\nu}_{n,\Re}^t = \frac{{\tilde v}^{t-1}_n}{2} - \frac{({\tilde v}_n^{t-1})^2}{2[(\sigma^2)^{t-1}+{\tilde v}^{t-1}_n]} \times \frac{1}{M}\sum\limits_{m=1}^{M}$\\ \ \ \ $\left[\frac{\eta^t_{m,n}\phi(\eta^t_{m,n})-\xi^t_{m,n}\phi(\xi^t_{m,n})}{\Phi(\eta^t_{m,n})-\Phi(\xi^t_{m,n})}
		+\left(\frac{\phi(\eta^t_{m,n})-\phi(\xi^t_{m,n})}{\Phi(\eta^t_{m,n})-\Phi(\xi^t_{m,n})}\right)^2\right]$;
		\STATE Set $\mathring{y}_{m,n} = \Im\{{\tilde y}_{m,n}\}, \mathring{u}^{t-1}_{m,n} = \Im\{{\tilde u}^{t-1}_{m,n}\}$, then calculate ${\hat z}_{m,n,\Im}^t$ and ${\nu}_{n,\Im}^t$ by repeating lines 5-8;
		\STATE ${\hat{\bf z}}_n^t = {\hat{\bf z}}_{n,\Re}^t + j{\hat{\bf z}}_{n,\Im}^t$, ${\nu}_n^t = {\nu}_{n,\Re}^t + {\nu}_{n, \Im}^t$;
		
		\vspace{0.5mm}
		\STATE $n\in {\cal L}:{\bf z}_{{\rm ext}, n}^t = \nu_{{\rm ext}, n}^t \left(\frac{{\hat{\bf  z}}_n^t}{{\nu}_n^t}-\frac{{\tilde {\bf u}}_n^t}{{\tilde v}_n^t}\right), \frac{1}{\nu_{{\rm ext},n}^t} = \frac{1}{{ \nu}^t_n}-\frac{1}{{\tilde v}_n^t}$; \\ \vspace{0.5mm} $n\in {\cal H}:{\bf z}_{{\rm ext}, n}^t = {\tilde {\bf y}}_{n}, \nu_{{\rm ext}, n}^t = (\sigma^2)^t$;
		\STATE $v_{n}^t = \frac{1}{M}\|{\bf{z}}^t_{\mathrm{ext},n}-{\bf{S}}{{\bf{u}}_{n}^{t-1}}\|^2_{2}- {\nu_{{\rm ext}, n}^t }$;
		\vspace{0.5mm}
		\STATE ${{\bf{r}}_{n}^{t}} = {{\bf{u}}_{n}^{t-1}} + \frac{K}{M}{{\bf{S}}^{\rm{H}}}\left({{\bf{z}}^t_{\mathrm{ext},n} - {\bf{S}}{{\bf{u}}_{n}^{t-1}}} \right)$;
		\vspace{0.5mm}
		\STATE ${\tau}_{n}^{t} = \frac{K-M}{M}{v}_{n}^{t} + \frac{K}{M}{\nu^t_{{\rm ext}, n}}$;
		\vspace{0.5mm}
		\STATE $({\pi^t_{k,n}})^{-1}=1+\frac{(1-\lambda^{t-1}_{k,n})(\tau^t_{n}+\psi^{t-1}_{n})}{\lambda^{t-1}_{k,n}\tau^t_{n}} \exp{(-\frac{\psi^{t-1}_{n}|r^t_{k,n}|^2}{\tau^t_{n}(\tau^t_{n}+\psi^{t-1}_{n})})}$;
		\STATE ${\mu}^t_{k,n}= {\pi}^t_{k,n} \frac{ {\psi^{t-1}_{n}}} {{\tau^{t}_{n}}+{\psi^{t-1}_{n}}} {r}^t_{k,n}$;
		\STATE${\gamma}^t_{n}= \frac{1}{K}\sum\limits_{k = 1}^K \left[{\pi}^t_{k,n} \frac{\tau^{t}_{n}\psi^{t-1}_{n}} {\tau^{t}_{n}+\psi^{t-1}_{n}} + (1-\pi^t_{k,n})|{\mu}^t_{k,n}|^2\right]$;
		\STATE $u_{k,n}^t = \frac{\tau_{n}^t {{\gamma}^t_{n}}}{\tau_{n}^t - {{\gamma}^t_{n}}}\left(\frac{\mu^t_{k,n}}{{\gamma}^t_{n}}- \frac{r_{k,n}^t}{\tau_{n}^t}\right)$;
		\STATE ${\psi}_{n}^{t} = {\sum\limits_{k=1}^{K}\left[\frac{\tau_{n}^{t-1}\psi_{n}^{t}}{\tau_{n}^{t}+\psi_{n}^{t-1}}+\left(\frac{\pi_{k,n}^{t} r_{k,n}^{t}}{\tau_{n}^{t}+\psi_{n}^{t-1}}\right)^2\right]}/{\sum\limits_{k=1}^{K}\pi_{k,n}^{t}}$;
		\STATE ${\lambda}_{k,1}^{t} = {\lambda}_{k,2}^{t} = \cdots = {\lambda}_{k, N_r}^{t} = \frac{1}{|{\cal M}|_c}\sum\limits_{n' \in {\cal M}}{\pi}_{k,n'}^{t}$;
		\STATE $(\sigma^2)^t = \frac{1}{|{\cal H}|_c}\sum\limits_{n \in {\cal H}} (\frac{1}{M}{{\|{\tilde{\bf Y}}_{:,n} - {\bf{S}}{\bm{\mu}_{n}^t}\|}_2^2}+ {{\gamma}^t_{n}})$;
		\vspace{0.5mm}
		\STATE${\tilde{\bf u}}^t_n = {\bf S}{\bf u}^t_n,({{\tilde {v}}_n^t})^{-1} = ({{{\gamma}^t_{n}}})^{-1}-({\tau_{n}^t})^{-1}$; 
		\ENDFOR
		\STATE Obtain ${\hat \alpha_k}$ according to (\ref{activity_detector}) ;
		\STATE \vspace{-0.5mm}
		\STATE {\textbf{\%\textit{Stage} 2}}:
		\STATE Initialize ${\bf Z}_{\mathrm{ext}}^{T_1 + 1} =  {\bf Z}_{\rm ext}^{T_1} {\bf D}_{\mathrm{sub}}$, ${\bf U}^{T_1} = {\bf U}^{T_1}{\bf D}_{\mathrm{sub}}$,\\
		\vspace{0.5mm} ${\bm \nu}_{{\rm ext}}^{T_1 + 1} = \text{diag}({\bf D}^{\mathrm H}_{\mathrm{sub}} {\text{diag}({\bm \nu}_{{\rm ext}}^{T_1})} {\bf D}_{\mathrm{sub}})$, $\lambda^{T_1+1}_{k,n} = \lambda^0_{k, n}$,\\
		\vspace{0.5mm} and ${\bm \psi}^{T_1 + 1} = \text{diag}({\bf D}^{\mathrm H}_{\mathrm{sub}} {\text{diag}({\bm \psi}^{T_1})} {\bf D}_{\mathrm{sub}})$;
		\vspace{0.5mm}
		\FOR {$ t=T_1+1$ {\textbf{to}} $ T_1 + T_2 $}
		\STATE $v_{n}^t = \frac{1}{M}\|{\bf{z}}^{T_1+1}_{\mathrm{ext},n}-{\bf{S}}{{\bf{u}}_{n}^{t-1}}\|^2_{2}- \nu_{{\rm ext}, n}^{T_1+1}$;
		\STATE Repeat lines 13-15 and let ${\pi^t_{k,n}} = {\pi^t_{k,n}} {\hat \alpha_k}$, then repeat lines 16-19;
		\STATE ${\lambda}_{k,n}^{t} =\frac{1}{|{\cal N}_n|_c}\sum\limits_{n' \in {\cal N}_n}{\pi}_{k,n'}^{t}$;
		\vspace{-1.5mm}
		\ENDFOR
		\vspace{0.5mm}
		\STATE $\hat{\bf X} = \left[{\bm \mu}_1^{T_1 + T_2},{\bm \mu}_2^{T_1 +T_2},\cdots,{\bm \mu}_{N_r}^{T_1 +T_2}\right]$.
		\vspace{-0.5mm}
	\end{algorithmic}
\end{algorithm}\hspace{-5mm}
variance is known \cite{JS_TSP,JS_TWC} or estimate it by neglecting the quantization noise \cite{KML_JSAC}, where the former is impractical and the latter is ineffective for low-bit quantized measurements.

When the iterative process terminates, the final posterior sparsity ratio is used to detect activity. The Bayesian activity 
detector is defined as follows:
\vspace{-1mm}
\begin{equation}\label{activity_detector}
	{\hat \alpha}_{k} = \begin{cases}
		1,&{\lambda}_{k,N_r}^{T_1} \geq 0.5,\\
		0,&\text{otherwise},
	\end{cases}
	\ \ \ \forall k.
\vspace{-1mm}
\end{equation}
If ${\hat \alpha}_{k} = 1$, the $k$-th device is considered to be active, otherwise it is inactive.

\subsubsection{Stage 2}
In this stage, our goal is to estimate the angular-domain channel from the intermediate results of \textit{Stage} 1. To mitigate the severe energy dispersion effect in near-field MIMO channels as illustrated in Fig. \ref{fig1}, we divide the whole XL-MIMO array into $N_{\rm sub}$ subarrays; then, the piecewise angular-domain channel can be represented as follows
\vspace{-1.5mm}
\begin{equation}\label{sub_angular_channel}
	{{\bf{X}}} = {{\bf{H}}}{{\bf{D}}_{\mathrm{sub}}},
\vspace{-1mm}
\end{equation}
where ${{\bf{D}}}_{\mathrm{sub}} = \text{Bdiag}({{\bf{D}}},\cdots,{{\bf{D}}}) \in {\mathbb C}^{N_r \times G_r}$ consists of $N_{\mathrm{sub}}$ traditional dictionary matrices ${\bf{D}} \in {\mathbb C}^{N_r/N_{\mathrm{sub}} \times G_r/N_{\mathrm{sub}}}$, and $G_r$ denotes the dimension of the dictionary. The $(m, n)$-th element of $\bf D$ is given by
\vspace{-1.5mm}
\begin{equation}\label{dict_matrix}
	{\bf{D}}_{m,n} = \frac{1}{\sqrt{G_r/N_{\mathrm{sub}}}} e^{j(m-1)\pi\frac{2(n-1)-G_r/N_{\mathrm{sub}}}{G_r/N_{\mathrm{sub}}}}. 
\vspace{-1mm}
\end{equation}
Accordingly, the equivalent SLM problem in (\ref{y_SLM_space}) becomes
\vspace{-1mm}
\begin{equation}\label{y_SLM_angular}
	{\bf{Z}}_{\rm ext}{{\bf{D}}}_{\mathrm{sub}}  = {\bf{S}}{{\bf{X}}} + {{\bf{W}_{\rm ext}}}{{\bf{D}}}_{\mathrm{sub}},
\vspace{-1.5mm}
\end{equation}
and we can adopt a similar approach as in \textit{Stage} 1 to estimate {{\bf{X}}}. It is worth mentioning that the activity pattern detected in \textit{Stage} 1 can be exploited as \textit{a priori} information to enhance the estimation performance.

The whole process of \textit{Stage} 2 is represented by lines 26-33 of \textbf{Algorithm 1}, where $T_2$ is the maximum number of iterations for this stage. Line 27 specifies the necessary transformation from the spatial-domain expression to the angular-domain expression. Lines 29-30 repeat the OAMP-MMV algorithm from \textit{Stage} 1, where the detected activity information is utilized to constrain $\pi^t_{k,n}$. The clustered sparsity structure of the angular-domain channel is adaptively learned in line 31, where ${\cal N}_n = \{n-1, n+1\} $  denotes the neighbor indices of $x_n$ in a subarray.

Similar to \cite{DAMP_AMP}, the \emph{damping} technique is adopted to effectively prevent the proposed TS-OAMP algorithm from divergence. Specifically, for line 18 of Algorithm 1, we have
%
${u}^t_{k,n} = (1-\epsilon){u}^t_{k,n} + \epsilon{u}^{t-1}_{k,n}$,
%
where the damping factor is chosen as $\epsilon = 0.4$. Likewise, ${\tilde v}^t_n$ is damped in the same way.

In Table I, we compare the computational complexity of Algorithm 1 with that of several baseline algorithms, including the structured sparsity-based generalized
approximated message passing (SS-GAMP) algorithm \cite{KML_JSAC}, the OAMP-MMV algorithm \cite{MYK_TWC}, and the simultaneous weighted orthogonal matching
pursuit (SWOMP) algorithm \cite{KML_TSP}. Here, we focus on the number of complex-valued multiplications and assume that the complexity of one real-valued multiplication equals one-quarter of the complexity of one complex-valued multiplication. Since the cubic term $N_rKM$ (or $G_rKM$) is the main contributor to the computational complexity, the number of complex-valued multiplications required for these algorithms is similar. Furthermore, the de-quantization processing of low-quantized signals inevitably causes additional computational complexity in the SS-GAMP, the OAMP-MMV, and the TS-OAMP algorithms.

\begin{table*}[!tp]
	\vspace{-6mm}
	\renewcommand\arraystretch{1.5}
	\caption{Complexity Analysis}
	\centering
	\begin{tabular}{c|c}
		\Xhline{1pt}
		Algorithm & Number of complex-valued multiplications\\
		\Xhline{1pt}
		SS-GAMP & $(T_1+T_2)\left(4N_rKM+20N_rK+16N_rM+3KM\right) + T_{\text{tur}}N_rM(K+1)$ \\
		OAMP-MMV & $(T_1+T_2)\left(\frac{37}{4}N_rK + 4G_rKM+G_rM+10G_rK+\frac{7}{4}G_r\right)$\\
		SWOMP & $(T_1+T_2)\left[N_rM(K+N_r)+(T_1+T_2+1)\left(N_rM+M\frac{2(T_1+T_2)+1}{3}+\frac{(T_1+T_2)(T_1+T_2+1)}{4}\right)\right]$ \\
		TS-OAMP & $T_1\left[N_r(4 K M + \frac{77}{4} K + M + \frac{3}{2})  + |{\cal H}|_c M (K+2)\right] + T_2G_r\left(3K M + 11 K+ M  + \frac{3}{4}\right)$ \\
		\Xhline{1pt}
	\end{tabular}
	\label{cc}
	\vspace{1mm}
	\\{Note: $T_{\text{tur}}$ denotes the iteration number of the SS-GAMP algorithm \cite{KML_JSAC}}.
	\vspace{-2mm}
\end{table*}

\vspace{-3mm}
\section{Simulation Results}
In this section, we evaluate the detection performance of the proposed TS-OAMP algorithm for near-field grant-free massive access.
Unless further specified, the main simulation parameters are set as follows: $N_r = 512$, $ K = 500$, $ K_a = 50$, $ \lambda = 0.05$ m, $ L_p = 5$, $B = 2$, $G_r / N_r = 2$, $N_{\mathrm{sub}} = 2$, $d^{\mathrm{LoS}}_{k,n} \in [10,100]$ m, $d^{\mathrm{NLoS}}_{k,n,l} \in (10,300)$ m, and $ \kappa_k = (13-0.03d^{\rm LoS}_{k, N_r/2})$ dB\footnotemark[2]$^{\text{,}}$\footnotemark[3].\footnotetext[2]{In this setup, the array length is around $12.8$ m, which is acceptable according to previous researches on XL-MIMO systems \cite{array_len_ref1, array_len_ref2}.}\footnotetext[3]{For $2$-bit quantization, we set $r_0 = -2.22$ and $ r_{2^B}=2.22$ in our experiments for computational simplicity and stability, avoiding the case that infinity times zero. In addition, $d^{\rm LoS}_{k, N_r/2}$ denotes the distance between the $k$-th device and the $\frac{N_r}{2}$-th antenna element at the BS. Since the distance between the device and the BS can vary dramatically for different antenna elements, we choose the middle position of the ULA as the representative for power control.}
The large-scale fading is modeled as $\beta_{k,n} = ({\lambda}/{4\pi d^{\rm LoS}_{k,n}})^2$, and adaptive power control is assumed at the devices, i.e., $P_k = P_t ({d^{\rm LoS}_{k, N_r/2}}/{100})^2$, where $P_t$ is the reference transmit power used by all devices for power control. 
Besides, the high-resolution ADCs are uniformly distributed in the extra-large array with an interval of $\Delta_{\cal H} = 32$, i.e., ${\cal H} = \{16,48,80,\cdots\}$.
The bandwidth is set to $1$ MHz and the power spectral density of the background noise is set to $-174$ dBm/Hz.
The maximum numbers of iterations of the proposed algorithm are set to $T_1 = 30$ and $T_2 = 20$.
We adopt the activity error rate (AER) and normalized mean square error (NMSE) as evaluation metrics
\vspace{-2mm}
\begin{equation}\label{AER_and_NMSE}
 {{\rm AER}} = \dfrac{1}{K}{{\sum\limits_{k = 1}^K {\left| {{{\hat \alpha }_k} - {\alpha _k}} \right|} }},\ {\mathrm{NMSE}} = {\mathbb E}\left[\frac{{\|{\bf H}-{\hat{\bf H}}\|}^2}{{\|{\bf H}\|}^2}\right],
\vspace{-1mm}
\end{equation}
where ${\mathbb E}[\cdot]$ denotes the expectation operator, and $\hat{\bf H}$ is the estimated spatial-domain channel. For the proposed algorithm, $\hat{\bf H}$ can be obtained based on $\hat {\bf X}$ and (\ref{sub_angular_channel}).
Additionally, we compare with the following baseline algorithms:

$\bullet$ {\textbf{SS-GAMP}}: The SS-GAMP algorithm \cite{KML_JSAC} exploits the common support structure to directly estimate the spatial-domain channel.

$\bullet$ {\textbf{OAMP-MMV}}: The OAMP-MMV algorithm \cite{MYK_TWC} utilizes the clustered sparsity structure and estimates the angular-domain channel, where the de-quantization module is added.

$\bullet$ {\textbf{SWOMP}}: The SWOMP algorithm \cite{KML_TSP} is a variant of the classic OMP algorithm. It also estimates the spatial-domain channel by exploiting the common support structure.
  
\begin{figure}[!t]
	\vspace{-0mm}
	\hspace{-4mm}
	\centering
	\subfigure[]{
		\includegraphics[width=4 cm, keepaspectratio]
		{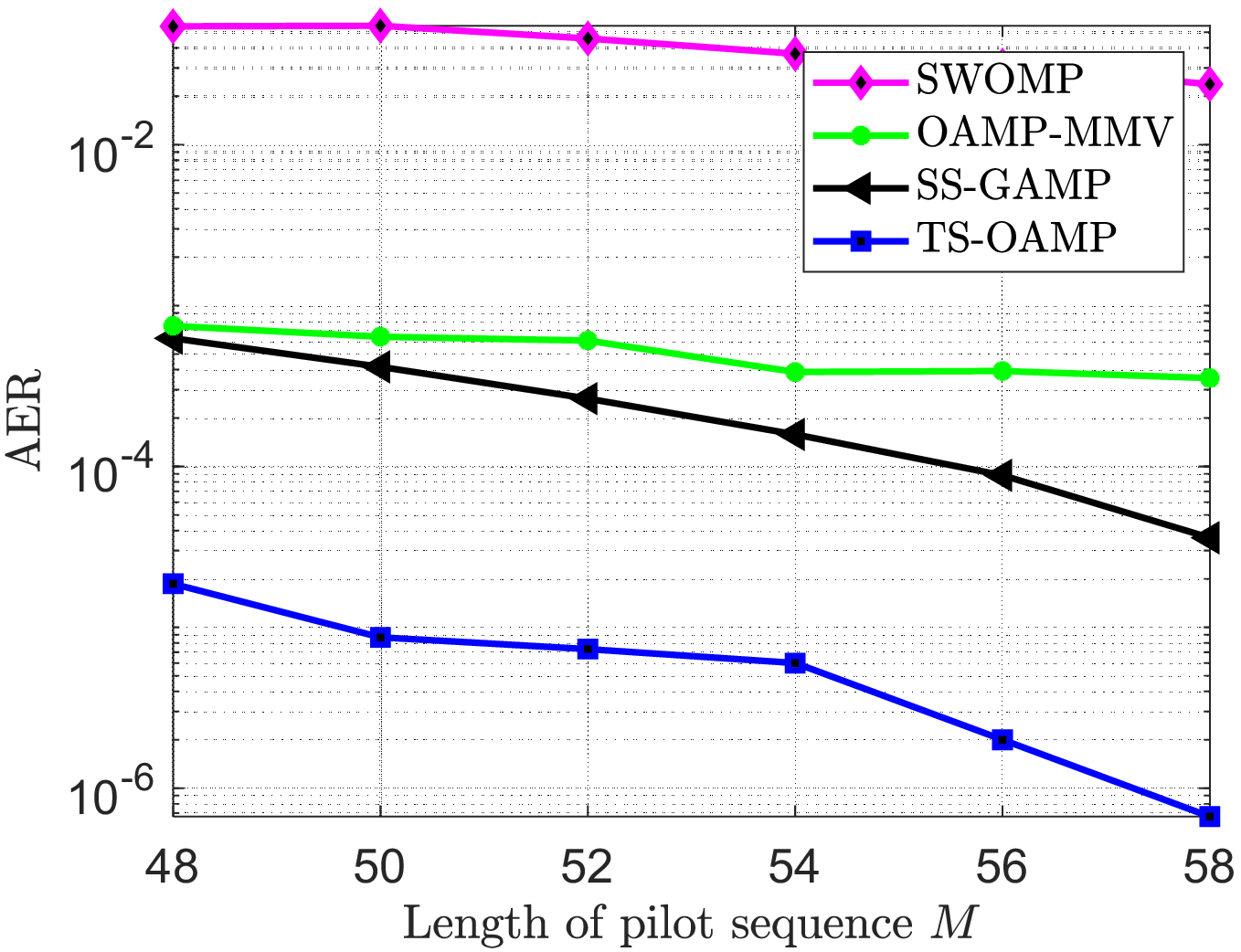}}
	\subfigure[]{
		\includegraphics[width=4 cm, keepaspectratio]
		{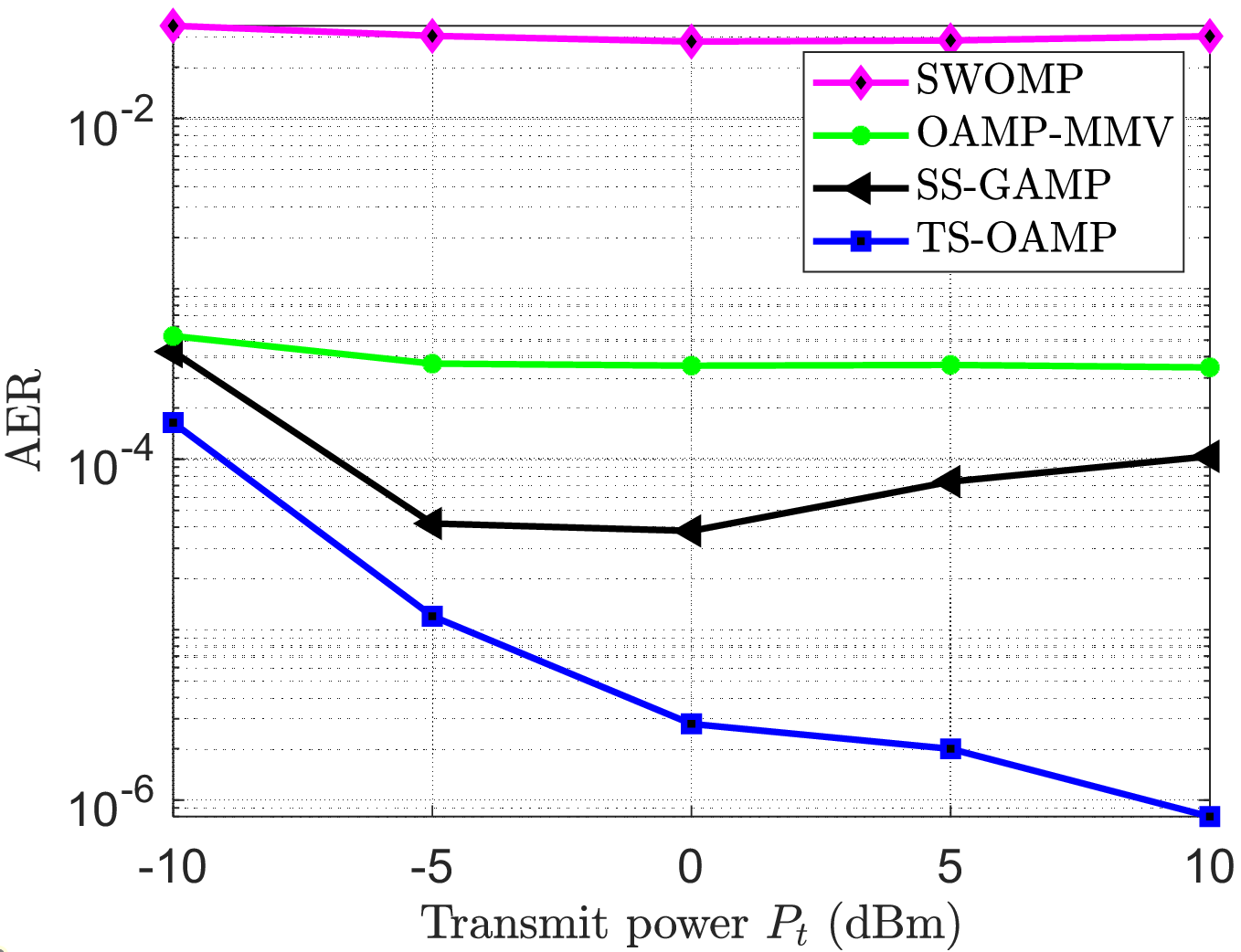}}
	\captionsetup{font={footnotesize}, singlelinecheck = off, justification = raggedright,name={Fig.},labelsep=period}
	\caption{AER of different JADCE algorithms versus: (a) the length of pilot sequence $M$ for $P_{t} = 5$ dBm; (b) the transmit power $P_t$ for $M=56$.}
	\label{Simulation_SNR}
	\vspace{-2mm}
\end{figure}

Fig. \ref{Simulation_SNR} shows the activity detection performance of different JADCE algorithms.
Although mixed-precision quantized signals are adopted for these detection algorithms, the proposed TS-OAMP algorithm achieves a high activity detection performance and outperforms the other algorithms due to the more efficient exploitation of the mixed-ADC architecture.

\begin{figure}[!t]
	\vspace{0mm}
	\hspace{-4mm}
	\centering
	\subfigure[]{
		\includegraphics[width=4 cm, keepaspectratio]
		{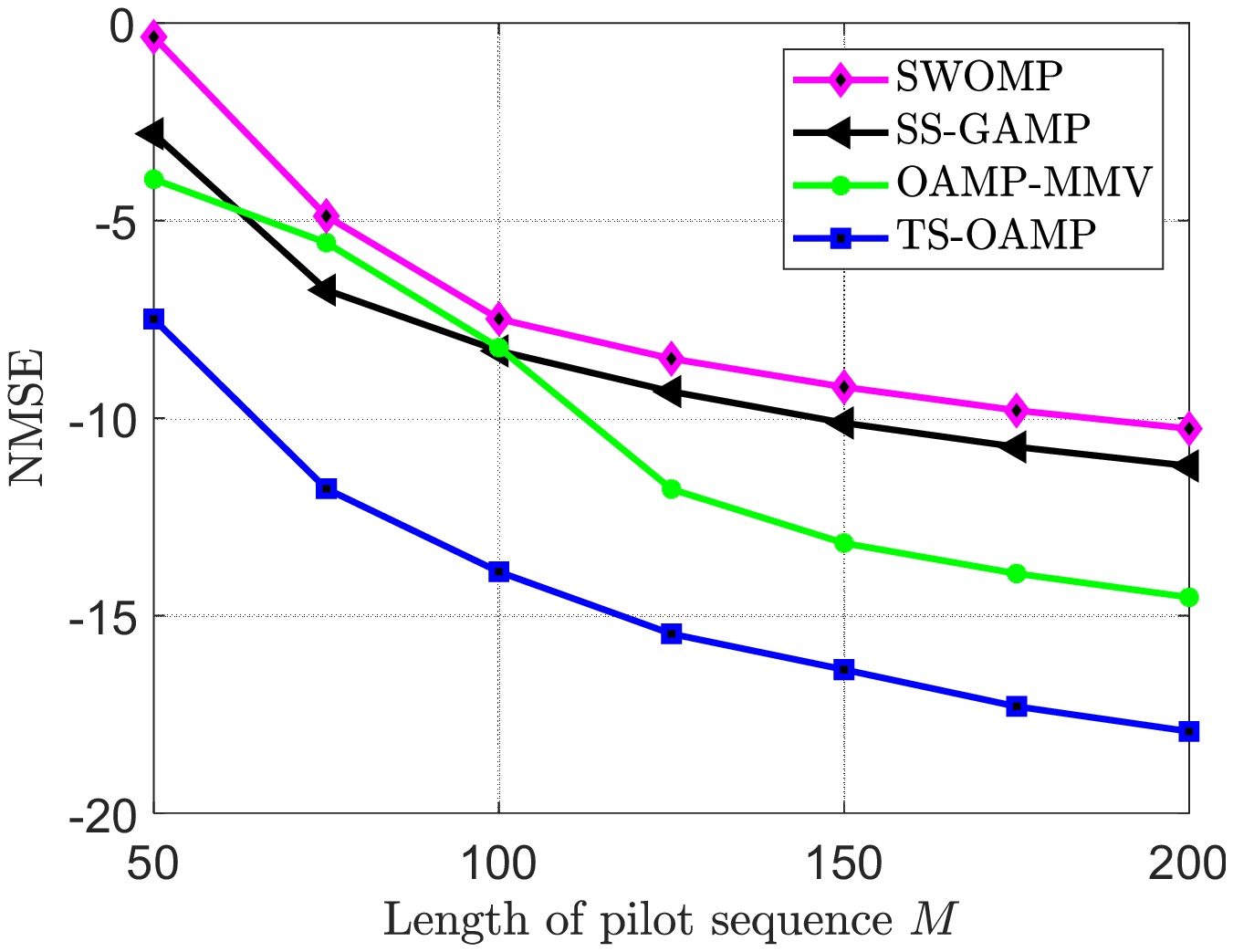}
	}
	\subfigure[]{
		\includegraphics[width=4 cm, keepaspectratio]
		{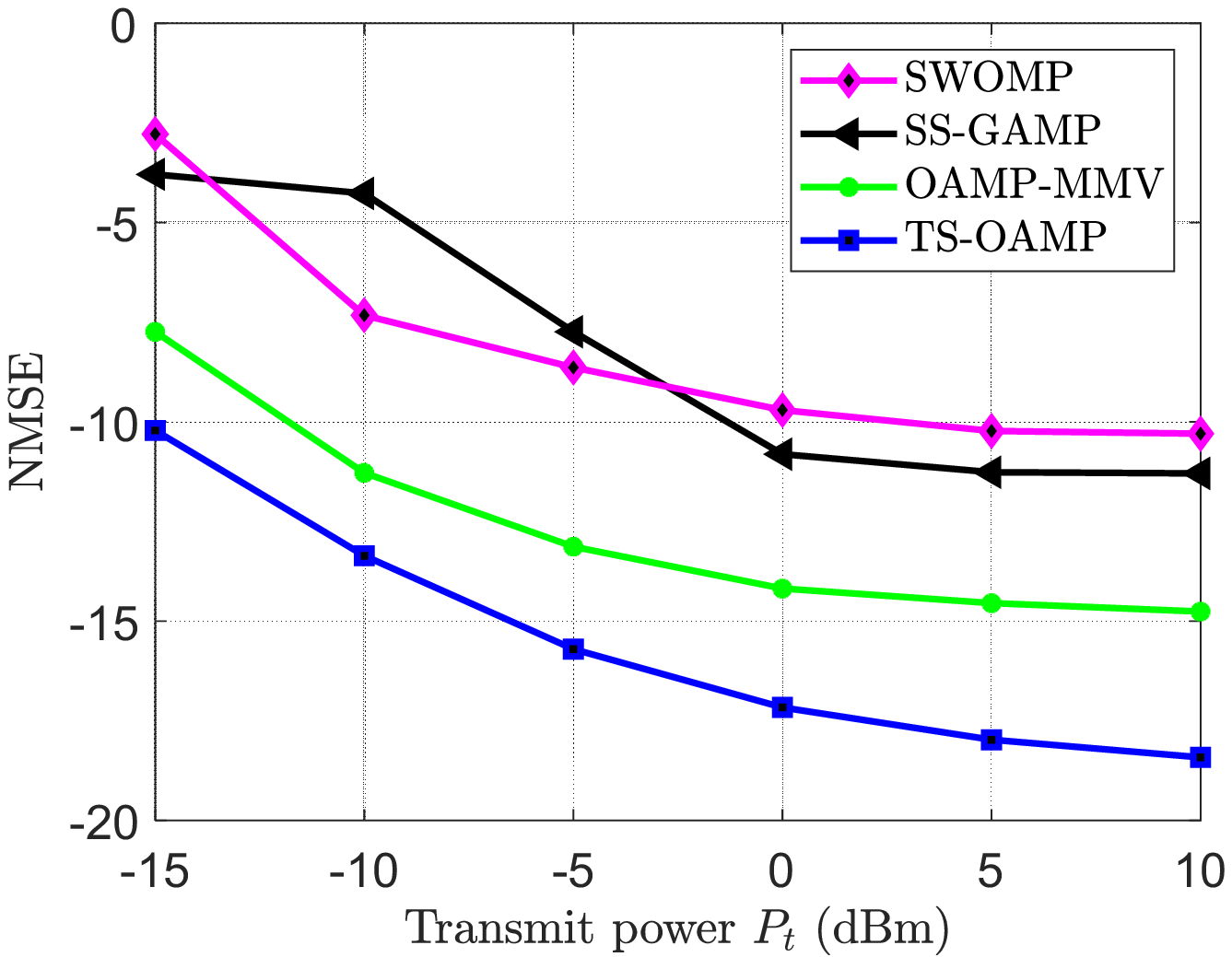}}
	\captionsetup{font={footnotesize}, singlelinecheck = off, justification = raggedright,name={Fig.},labelsep=period}
	\caption{NMSE of different JADCE algorithms versus: (a) the length of pilot sequence $M$ for $P_t = 5$ dBm; (b) the transmit power $P_t$ for $M=200$.}
	\label{Simulation_M}
	\vspace{-2mm}
\end{figure}

\begin{figure}[!t]
	\vspace{0mm}
	\hspace{-4mm}
	\centering
	\subfigure[]{
		\includegraphics[width=4 cm, keepaspectratio]
		{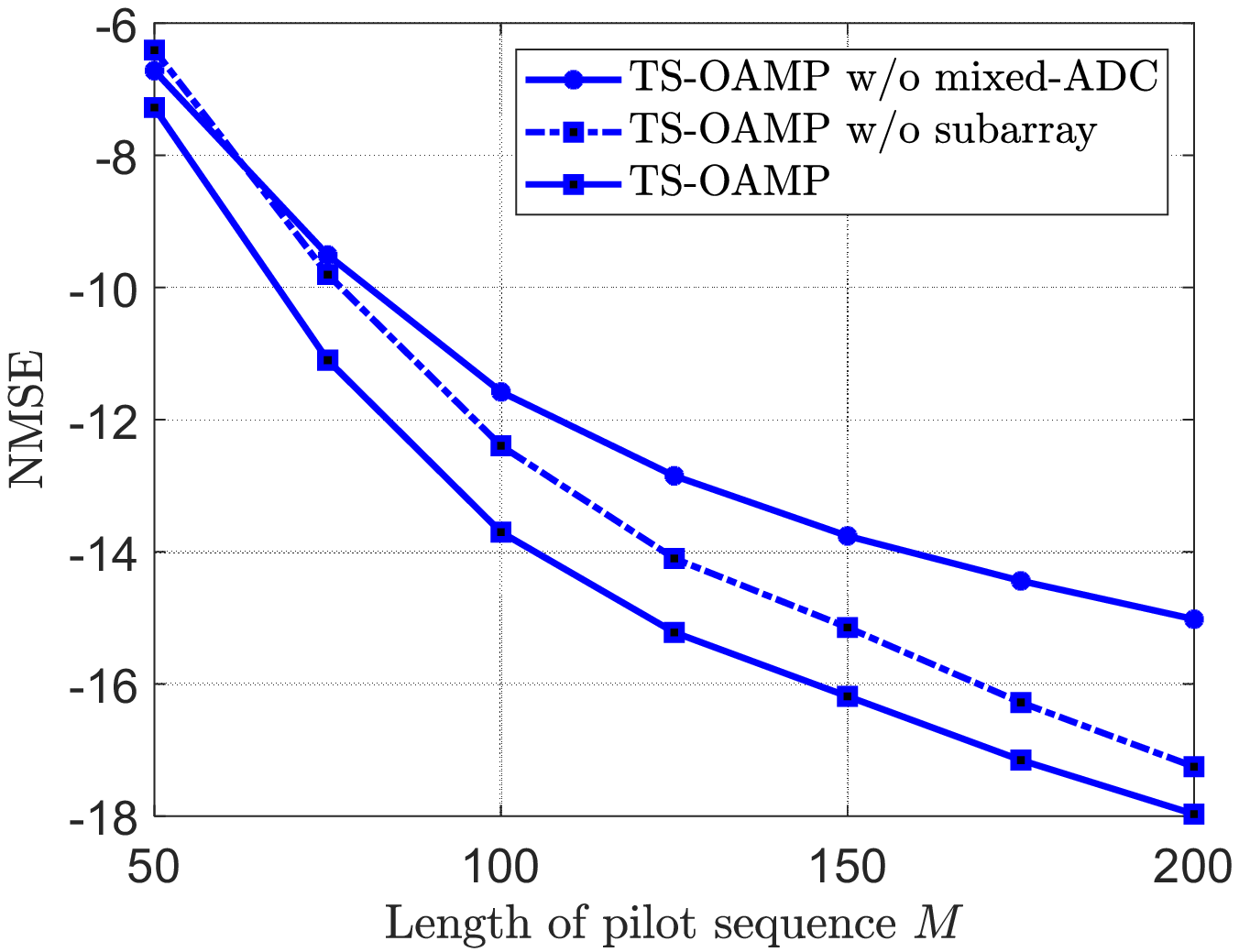}
	}
	\subfigure[]{
		\includegraphics[width=4 cm, keepaspectratio]
		{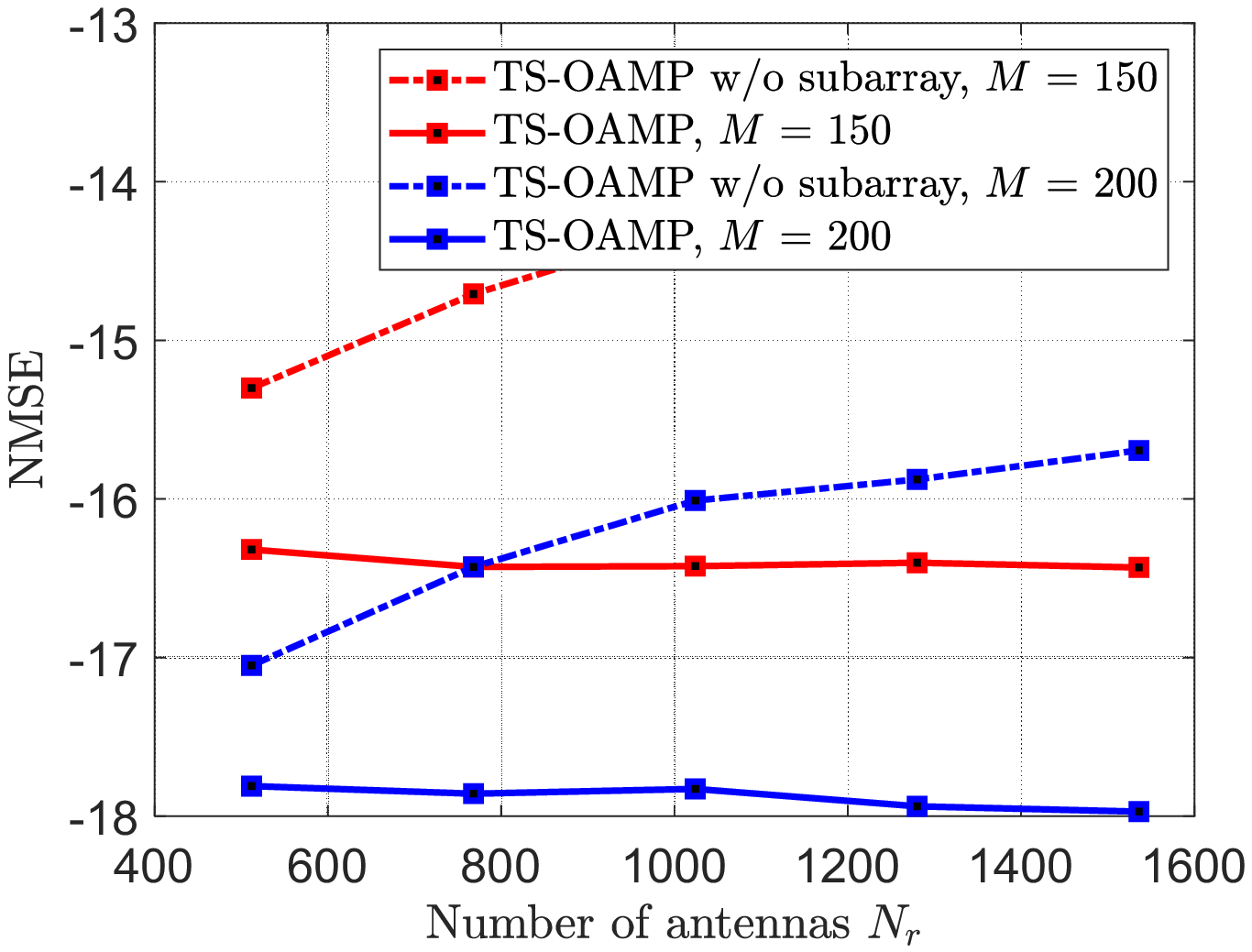}}
	\captionsetup{font={footnotesize}, singlelinecheck = off, justification = raggedright,name={Fig.},labelsep=period}
	\caption{(a) Gains of the proposed algorithm for the mixed-ADC architecture and subarray-wise estimation versus $M$; (b) Gains of the proposed algorithm with subarray-wise estimation versus $N_r$.}
	\label{Simulation_Gain}
	\vspace{-4mm}
\end{figure}

In Fig. \ref{Simulation_M}, we compare the NMSE performance of different JADCE algorithms. As can be seen, the OAMP-MMV algorithm and the TS-OAMP algorithm perform better since they exploit the sparsity of the angular-domain channel. However, the proposed algorithm significantly outperforms the OAMP-MMV algorithm thanks to the subarray-wise processing and more efficient utilization of the mixed-ADC architecture.

\begin{figure}[!t]
	\vspace{-3mm}
	\hspace{-4mm}
	\centering
	\includegraphics[width=4 cm, keepaspectratio]
	{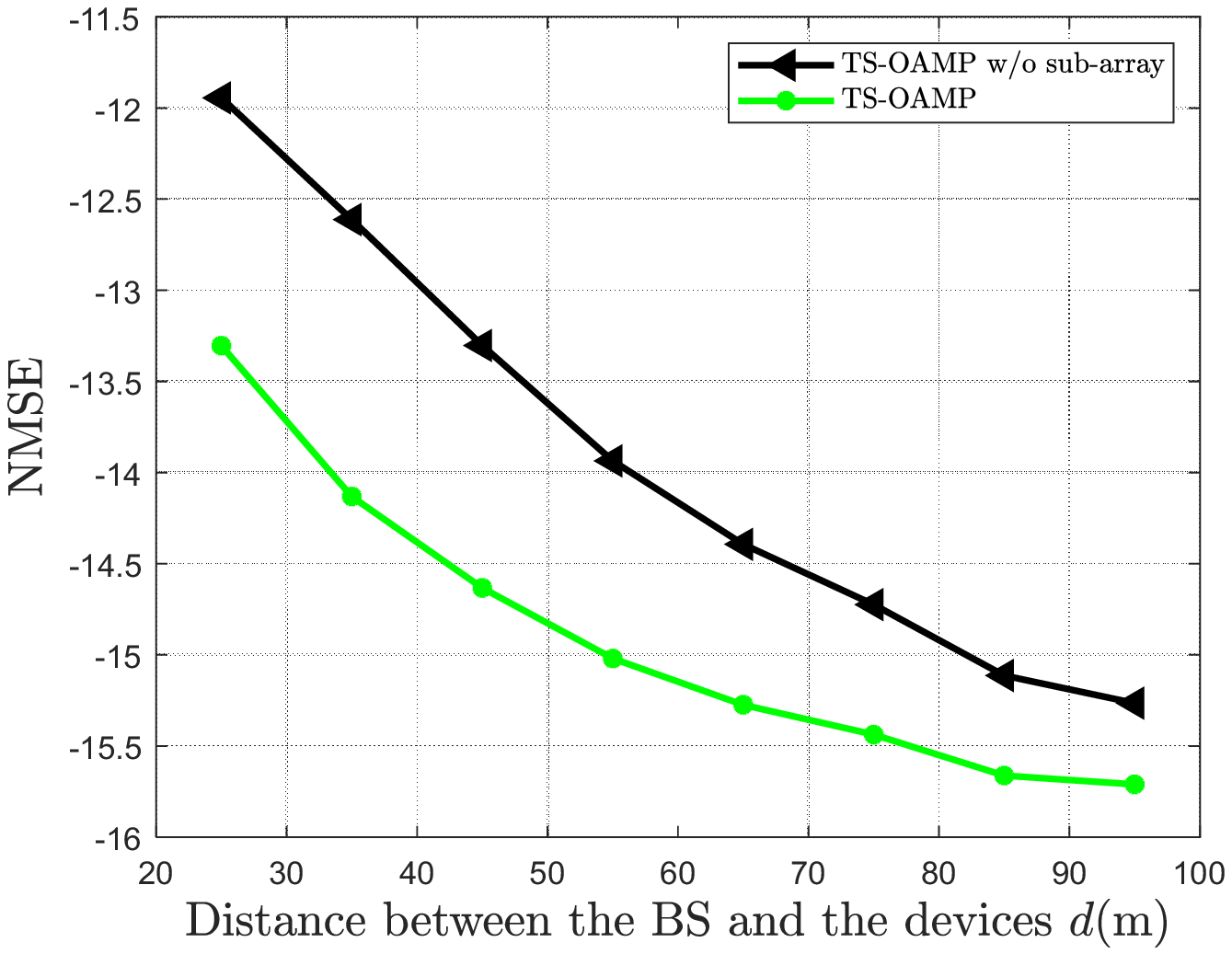}
	\captionsetup{font={footnotesize}, singlelinecheck = off, justification = raggedright,name={Fig.},labelsep=period}
	\caption{NMSE performance of the proposed algorithm versus the distance between the BS and the devices, $d$, for $M = 100$ and $P_t = 10$ dBm.}
	\label{Simulation_Distance}
	\vspace{-2mm}
\end{figure}

\begin{figure}[!t]
	\hspace{-4.5mm}
	\centering
	\subfigure[]{
		\includegraphics[width=4 cm, keepaspectratio]
		{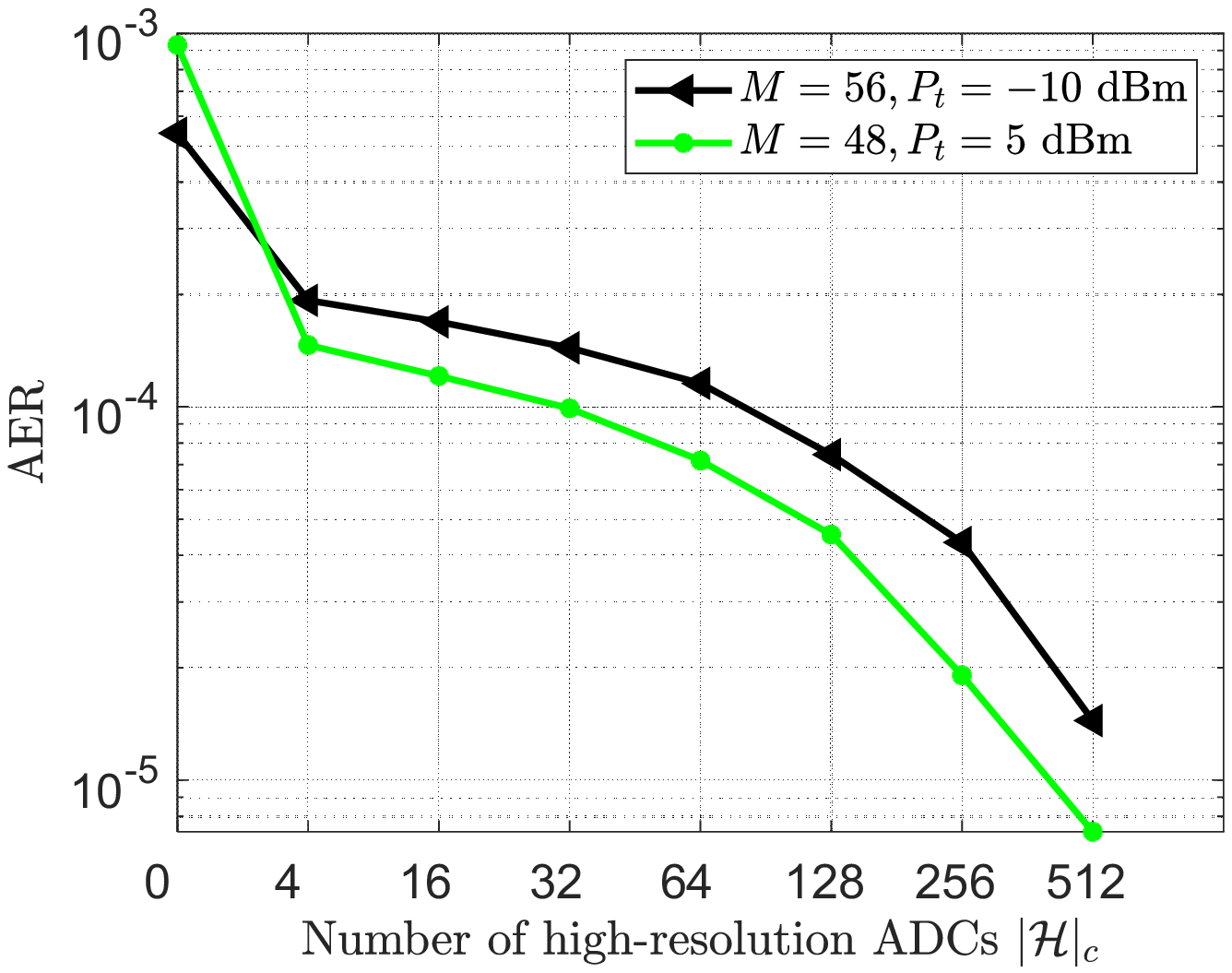}
	}
	\subfigure[]{
		\includegraphics[width=4 cm, keepaspectratio]
		{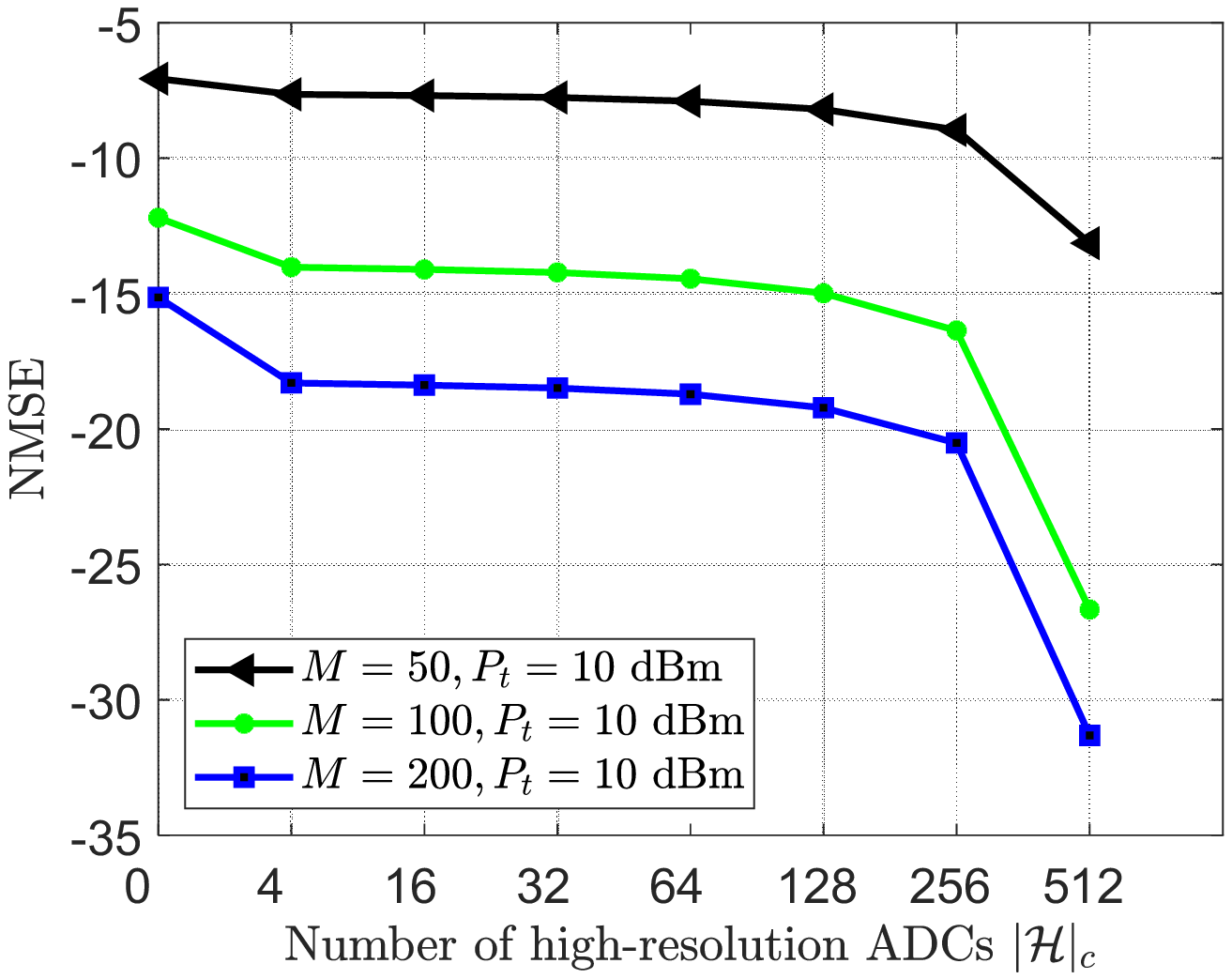}}
	\captionsetup{font={footnotesize}, singlelinecheck = off, justification = raggedright,name={Fig.},labelsep=period}
	\caption{Detection performance of the TS-OAMP algorithm versus the number of high-resolution ADCs $|{\cal H}|_c$: (a) AER; (b) NMSE.}
	\label{Simulation_ADCs}
	\vspace{-4mm}
\end{figure}

Fig. \ref{Simulation_Gain} verifies the gain of the proposed algorithm for the mixed-ADC architecture and for subarray-wise estimation. Without the assistance of high-resolution ADCs, the detection performance of the TS-OAMP algorithm suffers even if the two-stage structure and subarray-wise estimation are employed, as can be observed in Fig. \ref{Simulation_Gain}(a). Besides, we can observe the benefits of subarray-wise processing, demonstrating its effectiveness in suppressing the near-field effects. 

The impact of distance on the NMSE performance of the proposed algorithm is studied in Fig. 6, where the devices are distributed in the range $d^{\mathrm{LoS}}_{k,n} \in (d-5,d+5)$ m. As can be seen, the benefit of subarray-wise processing decreases with distance, since the near-field effect is more pronounced when the devices are closer to the BS. As the distance increases, the detection performance of both the TS-OAMP algorithm and the TS-OAMP algorithm without subarray-processing improves due to the weakening of the near-field effect.



In Fig. 7, we investigate the detection performance of the TS-OAMP algorithm as a function of the number of high-resolution ADCs $|{\cal H}|_c$ in the mixed-ADC architecture. Generally, the detection performance of the TS-OAMP algorithm improves as $|{\cal H}|_c$ increases, and the TS-OAMP algorithm employing only high-resolution ADCs and only low-resolution ADCs has the best and worst detection performance, respectively. Moreover, Fig. 7 shows that a small number of high-resolution ADCs can significantly enhance the detection performance compared to the case with only low-resolution ADCs, which verifies the effectiveness of the mixed-ADC architecture in our proposed scheme. 

\vspace{-3mm} 
\section{Conclusions}
In this paper, we proposed a TS-OAMP algorithm to solve the JADCE problem for massive connectivity based on mixed-ADC XL-MIMO over near-field channels. In the first stage, the activity detection relied on the common support structure of spatial-domain channels and the accurate hyper-parameter estimation from high-precision measurements. Next, the sparse angular-domain channel was estimated based on the intermediate results of the previous stage, where the near-field effect of the XL-MIMO channels was mitigated by subarray-wise processing. Numerical results verified that our proposed algorithm performs remarkably well facilitating near-field massive access in XL-MIMO systems. In future work, the proposed schemes can be extended to multi-carrier systems to enhance the range of possible applications.

\end{document}